# A Dynamic Framework of Reputation Systems for an Agent Mediated e-market

**Vibha Gaur[1], Neeraj Kumar Sharma[2]**

**[1] Department of Computer Science, University of Delhi**
**Delhi, India**
***3.vibha@gmail.com***

**[2] Department of Computer Science, University of Delhi**
**Delhi, India**
***neerajraj100@gmail.com***

**Abstract**

The success of an agent mediated e-market system lies in the underlying reputation management system to improve the quality of services in an information asymmetric e-market. Reputation provides an operatable metric for establishing trustworthiness between mutually unknown online entities. Reputation systems encourage honest behaviour and discourage malicious behaviour of participating agents in the e-market. A dynamic reputation model would provide virtually instantaneous knowledge about the changing e-market environment and would utilise Internets' capacity for continuous interactivity for reputation computation. This paper proposes a dynamic reputation framework using reinforcement learning and fuzzy set theory that ensures judicious use of information sharing for inter-agent cooperation. This framework is sensitive to the changing parameters of e-market like the value of transaction and the varying experience of agents with the purpose of improving inbuilt defense mechanism of the reputation system against various attacks so that e-market reaches an equilibrium state and dishonest agents are weeded out of the market.

## 1. Introduction

With the growing popularity of e-commerce and amount of information on WEB, users expect automated techniques to assure the trustworthiness of information available on internet. Software agents offer a promise to change e-commerce trading by helping internet traders to purchase products from online distributed resources based on their interests and preferences [16]. Assuring the trustworthiness of web products and services in such an environment where actual traders may never meet each other is a challenging task performed by reputation systems. Reputation systems have a high utility in those environments where entities are long lived, feedback about the current interactions is captured and distributed, and past feedback/experience guides buyer decisions [22]. These systems are oriented to develop trustworthiness or the degree to which one agent has confidence in another within the context of a given purpose or decision.

The definition and meaning of reputation varies with applications and contexts. From an objective view, reputation is expressed as "a quantity derived from the underlying social network which is globally visible to all members of the network" [25] or, "a perception that an agent has of another's intentions and norms" [17]. Reputation and Trust are often used in complementary fashion as an agent expects positive outcomes when interacting with another agent that has a reputation for being trustworthy [8]. Some systems are described as trust systems as therein agents determine whether another agent will do what it says it will, whereas others are best described as reputation systems because therein agents compute and propagate their beliefs about other agents.

The e-market environment in which these agents operate is generally open, that means agents can join or leave the marketplace at any time; uncertain, i.e. the true worth of a good can be judged only after its purchase; and un-trusted, that is the e-market comprises of honest and dishonest agents. The e-market is populated with self interested buyer and seller agents that try to maximise their respective gains. The e-market environment is itself dynamic in nature as it undergoes continuous changes with different agents joining and leaving the e-market at will. The power of a reputation system in an agent mediated e-commerce can be realized to the optimum if different process models inherent to the e-transactions like deciding about pricing of goods, computing and distributing reputation of participants and selection of a seller for purchasing a good are also dynamic [29]. A truly dynamic model must be sensitive to the changing e-market environment and must adapt to changing experience of buyer/seller agents with each transaction. Dynamic e-market models would provide virtually instantaneous knowledge about the changing e-market environment and would utilise Internets' capacity for continuous interactivity. Designing efficient and robust reputation systems that satisfy both the buyers as well as sellers is a challenge for the research community.

The objective of this paper is to propose a framework for a dynamic reputation system that is sensitive to the changing parameters of the dynamic e-market environment like the experience of agents involved in transactions, value of a transaction and number of transactions between the same buyer-seller pair. In the proposed model, each of the individual process model of the dynamic reputation framework is itself dynamic as selection of a seller for buying a good depends on the changing experience between a buyer-seller pair; computing sellers' reputation by a buyer depend on the experience of an agent in the e-market, mutual experience of a buyer-seller pair and the value of transaction. Further, incorporating value of transaction in reputation computation affects the amount of reputation that is to be enhanced or reduced after each transaction. This makes the reward/penalty proportional to the size of the transaction in which honest/dishonest behavior is exhibited by seller agents, and negates any benefit of a Value Imbalance attack where a seller agent gains reputation by showing honesty for small value transactions and then cheats for a large value transaction. Making the reputation updation dependent on the experience of agents, by varying the weightage of individual experience and shared opinion from others, reduces the effect of Ballot Stuffing attack where a number of malicious agents artificially enhance or reduce the reputation of another agent. Also, by making the reputation updation sensitive to the fact that whether reputation is earned from a single buyer or multiple buyers minimizes the effect of collusion between a buyer-seller pair. The proposed framework employs judicious use of information sharing and thus reduces the associated cost by using effective inter-agent communication.

The reputation computation strategy proposed in this paper uses reinforcement learning (RL) techniques which provide a general framework for sequential decision making problems [10]. RL deals with what an agent should do in every state that it can be and how to map situations to action, in order to maximize the long term reward. The learner must discover which actions yield the maximum reward by trying them. Sometimes, actions may affect not only the immediate reward, but also all subsequent rewards. Hence, trial-and-error search and delayed reward are the two most important distinguishing features of RL.

In the proposed strategy, for purchasing a good, the buyer chooses a seller offering the highest expected value of the good i.e. good with highest expected utility for the buyer. Expected buyers' requirement from a good constitutes buyers' estimation of goods' attributes and is subjective and fuzzy in nature. It is subjective as relative priority of attributes of a good would vary with each good and with each buyer. It is fuzzy as generally buyers' expectations of a particular attribute are specified in fuzzy terms like "low" or "high". Similarly, a buyer has to map linguistic assessment of goods being offered by different sellers based on their attributes to the fuzzy scale. Hence this paper uses fuzzy set theory to allow a buyer agent to compute attribute weights of a good and to select a seller that offers the good with highest expected value.

The rest of this paper is organized as follows. Various reputation models from literature and in commercial use are introduced in section 2. Section 3 presents the proposed dynamic reputation framework. To address existing problems, section 4 illustrates the performance of the proposed system against known attacks. A case study is presented in section 5. Section 6 concludes the paper.

## 2. Related Work

Reputation models are an important component of e-market, help building trust and elicit cooperation among loosely connected and geographically dispersed economic agents [12]. A number of reputation models described in literature are discussed below.

The evidential model [2, 3] for reputation computation assumes a distributed reputation environment and is based on Dampster Shafer Theory. An agent finds the trustworthiness of another agent [3] based on its direct interaction and testimonies of other trustworthy agents.

Some reputation models [21, 26] from literature employ reinforcement learning and are based on individual experience only. In reputation model for increasing user satisfaction [26], seller agents adjust the price and quality of goods to maximise their profit. A multi-facet reputation model [21] involves reputation computation of both buyer and seller agents using quality, price and delivery time of goods. But, these systems [21, 26] suffer heavily from re-entry and multi-identity attacks as these use negative reputation and new sellers do not start from minimum reputation.

TRAVOS [15] employs Bayesian probability analysis and computes trust of an agent by taking into account past experience between two agents, and in case of lack of past experience, this model utilizes the information collected from third parties. To filter out unfair opinions, TRAVOS uses an endogenous approach to filter out unfair opinions.

PeerTrust [19] is a reputation model that uses techniques for resilient reputation management against vulnerabilities like feedback sparsity and feedback manipulation. It talks about dynamism in electronic communities from the perspective of honest and dishonest behaviour of actors.

Reputation in Gregarious societies (REGRET) [16, 17] employs fuzzy rules to find reliability of witness agents based on their relationship with the target agent. REGRET is a multi-facet reputation mechanism that models the reliability of reputation based on the number of interactions of witness agents with the target agent.

Another model "Trunits" [24] is based on accumulation of trust units (trunits). A seller must possess sufficient number of trunits before executing a transaction. To engage in a transaction, seller must risk a particular quantity of trunits which is put into an escrow with the market operator. After a transaction, if buyer is satisfied, seller gets more trunits, otherwise it loses risked ones.

Broker assisting TRS [4] is a flexible model based on Artificial Neural Networks (ANN) that employs backpropagation algorithm. Use of ANN helps to reduce noise data and supports incremental training, so each agent requests for information only from those having a similar reputation evaluation criterion.

In Reputation Dynamics and Convergence [8], authors formalize the desiderata that from a dynamic systems' perspective a reputation system should have the properties of Monotonicity and Accuracy. As an example of Monotonicity, agents who provide high quality goods at low price should acquire better reputation and, in systems with focus on Accuracy, the buyer should quickly learn the accurate reputation value for the seller. The Dynamic Framework proposed in this paper incorporates Monotonicity as the process of seller selection and also updation of reputation are based on the presence of favourable goods' attribute like low price and high quality. Further, a fraudulent seller is penalised immediately to keep the reputation estimate accurate.

The P4P (Pervasive Platform for Privacy Preferences) [20] system concentrates on privacy control in case of e-transactions. The paper acknowledges the property of e-market environment being dynamic and, the need that the existing systems in this environment should also be dynamic. It emphasizes importance of reputation by allowing the clients, the freedom to not disclose personal data according to the level of reputation.

A number of simple online reputation systems are in commercial use. eBay [14] is the most popular auction site that has feedback forum as a reputation system in which after each transaction, a buyer rates a seller as positive, negative or neutral i.e. +1, -1 or 0 respectively. The reputation of a user is computed by subtracting total number of negative feedbacks from the total number of positive feedbacks obtained from distinct users [23]. Amazon [13] is America's largest online retailer where reviews include star ratings from 1 to 5 and a prose text. Average of all ratings is used to assign reputation.

A limitation of the existing systems from literature [1, 3, 4, 8, 15, 16, 19, 20, 21] is that, during the process of computing or updating of reputation values, these do not take into consideration the changing parameters of dynamic e-market environment like the varying experience of agents and the value of a transaction which make them vulnerable to different attacks. The proposed reputation framework incorporates value of a transaction in the strategy of reputation computation to remove the problem of Value Imbalance attack and, varies the weightage of individual and shared reputation components with changing experience of agents to minimise the effect of Ballot Stuffing attack.

## 3. Dynamic Reputation System Framework

Reputation systems are oriented to encourage trustworthy behaviour, increase user satisfaction and deter dishonest participants by providing means through which reputation could be computed and disseminated [22]. The e-market environment in which reputation systems operate is dynamic as it changes continuously in terms of agents freely entering/exiting the market and also with the varying experience of agents. Therefore, as a buyer gains experience of a sellers' behaviour with each repeated transaction, the weightage of the individual experience of a buyer-seller pair should increase as compared to the opinion shared by other buyer agents. Moreover, economic worth of being honest or dishonest in a transaction cannot be judged without taking into account the value of a transaction as honest behaviour in a large transaction is more important than in a small transaction.

A dynamic reputation framework should base the reputation computation methodology itself on the dynamics of the e-market environment to infuse some inbuilt defense capability against possible attacks. In order to have a robust and high utility reputation system, different activities belonging to reputation computation methodology it should be adaptive to the changing environment and the experience of agents involved in a transaction. The next section describes the proposed dynamic reputation computation strategy that employs reinforcement learning and fuzzy set theory.

## 3.1 Buyers' Strategy for Reputation Computation

The proposed buyers' strategy is based on the e-market model having a set of buyers and sellers. In this model, sellers are divided into four categories, namely, reputed, non-reputed, dis-reputed and new sellers. The reputation of seller *s* being computed by buyer *b* is composed of two components: individual reputation and shared reputation. These two components are combined to represent overall reputation of a seller agent.

In this model, B represents the set of buyers, S represents the set of sellers and, G the set of goods. Let $r_t^b \in [0,1)$ represents individual reputation (IR) component, $or_t^{others} \in [0,1)$ represents shared reputation (SR) i.e. the opinion of other buyers for seller *s*, and $or_t^b \in [0,1)$ represents overall reputation of seller *s* at time t, for the buyer *b*. At time t+1, buyer *b* stores/remembers the overall reputation $or_t^b(s_i)$ of all sellers $s_i \in S$, with whom buyer *b* has interacted at time t in the past. Each buyer maintains four categories of sellers as defined below.

(i) $S_R^b$ : Sellers in the reputed list of buyer *b*, i.e. $or^b(s) \geq \Theta^b$, where $s \in S_R^b$ , $\Theta^b$ is the reputation threshold of buyer *b* and $0 < \Theta^b < 1$ .

(ii) $S_{NR}^b$ : Sellers in the non-reputed list of buyer *b*, i.e. $\theta^b < or^b(s) < \Theta^b$ where $s \in S_{NR}^b$ .

(iii) $S_{DR}^b$ : Sellers in the dis-reputed list of buyer *b*, i.e. $0 < or^b(s) \leq \theta^b$ , where $s \in S_{DR}^b$ , $\theta^b$ is the dis-reputation threshold and $0 < \theta^b < \Theta^b$.

(iv) $S_{NewR}^b$ : Sellers that are new to buyer *b* in the market, initially $or^b(s) = 0$. A new seller s remains in this list until its reputation crosses the dis-reputation threshold $\theta^b$. Before crossing $\theta^b$, if a seller cheats than it is moved to the list of dis-reputed sellers $S_{DR}^b$ and is never considered again for business.

The process of choosing a seller for purchasing a good based on its expected value uses three important algebraic operations on fuzzy numbers: inverse, addition and multiplication. If $\tilde{A} = (a_1, a_2, a_3, a_4)$ and $\tilde{B} = (b_1, b_2, b_3, b_4)$ are two positive trapezoidal fuzzy numbers then, the fuzzy addition of $\tilde{A}$ and $\tilde{B}$ is given in (1) and inverse of a fuzzy number $\tilde{A}$ represented as $\tilde{A}^{-1}$ is shown in (2) below.

$$\tilde{A} + \tilde{B} = (a_1 + b_1, a_2 + b_2, a_3 + b_3, a_4 + b_4) \quad (1)$$

$$\tilde{A}^{-1} = \left(\frac{1}{a_4}, \frac{1}{a_3}, \frac{1}{a_2}, \frac{1}{a_1}\right) \quad (2)$$

Unlike addition and subtraction, product of two trapezoidal fuzzy numbers may not result into a trapezoidal number [6, 7]. Therefore, this paper uses an approximation of the product of two trapezoidal fuzzy numbers to a new trapezoidal fuzzy number [7]. The product of two trapezoidal fuzzy numbers, A and B given above is approximated by the trapezoidal fuzzy number $\tilde{C} = \tilde{A} \times \tilde{B} = (c_1, c_2, c_3, c_4)$ as proposed in [6, 7] where,

$$c_1 = \frac{3}{2}(a_2 - a_1)(b_2 - b_1) + 2\begin{bmatrix}(a_2 - a_1)b_1 \\ +(b_2 - b_1)a_1\end{bmatrix} + 3a_1b_1 - 2a_2b_2 ,$$

$$c_2 = a_2b_2,$$

$$c_3 = a_3b_3 ,$$

$$c_4 = \frac{3}{2}(a_4 - a_3)(b_4 - b_3) - 2\begin{bmatrix}(a_4 - a_3)b_4 \\ +(b_4 - b_3)a_4\end{bmatrix} + 3a_4b_4 - 2a_3b_3 \quad (3)$$

For defuzzifying, Centre Of Area (COA) or Centroid method is used. For a fuzzy number $\tilde{A} = (a_1, a_2, a_3, a_4)$, its COA is computed as: $(a_1 + a_2 + a_3 + a_4)/4$.

The proposed reputation computation methodology based on the concept of reinforcement learning and fuzzy set theory is divided into three phases. In Phase I, a buyer expresses its willingness to buy a good and the set of sellers' who respond for selling that good are elicited and a seller selection methodology using fuzzy arithmetic is applied to select a seller for purchasing that good. Phase II includes reputation computation using reinforcement learning. It begins after purchasing the good, where the buyer updates the sellers' reputation based on the experience of the current transaction and the opinion from others. Finally in Phase III, the buyer updates its list of reputed, non-reputed, dis-reputed and new sellers. A detailed description of this methodology divided into Phase I, Phase II and Phase III is given below.

Phase I:

1. The process of buying and selling starts with a buyer *b* announcing the need to buy a good *g* by sending broadcast request to all sellers. Those sellers who are willing to sell good *g* respond by submitting their bids. At any given time, buyer *b* preferably purchases a good from a reputed seller. If no seller from the reputed list offers good *g* then the buyer *b* selects a seller from the set of non-reputed sellers but in no case the buyer would choose a dis-reputed seller [27]. In addition, with a small probability ρ, buyer *b* would choose a seller from the list of new sellers' i.e. $S_{NewR}^b$ . Initially the value of ρ is 1 and it decreases over time to some minimum value defined by buyer *b*.

2. After receiving sellers' bids for good *g*, buyer *b* first computes the expected value $f^b(g, s)$ of good g's offer from each seller and then selects a seller s that is offering good g with highest expected value i.e. max. $f^b(g, s)$ based on the following strategy by computing goods' attribute weights using extent analysis method [5, 28] and combining it with fuzzy AHP technique.

   i. Obtain the buyers' assessment of pairwise comparison of different attributes of a good in linguistic terms like Equally important (E), Moderately Important (M), Highly Important (H), Very Highly Important (VH) or Extremely Important (EI) as illustrated in Fig. 1.

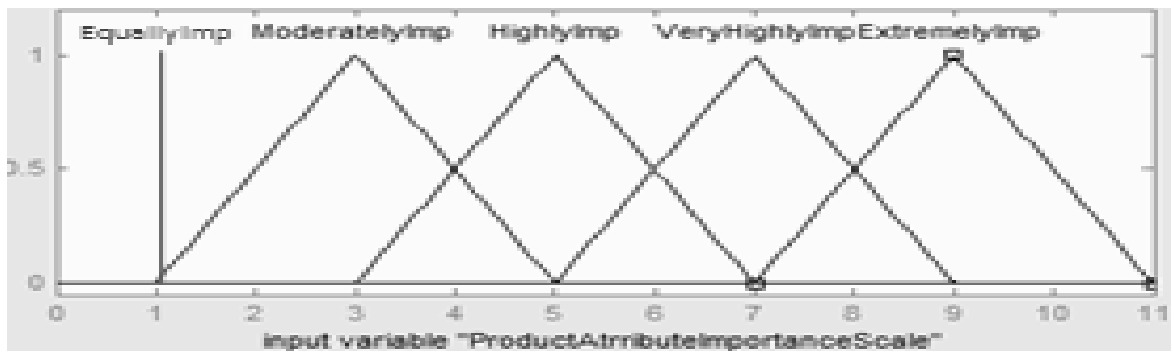

Fig. 1 Fuzzy Scale for Relative Importance of Attributes

Using fuzzy scale of Fig. 1, map these linguistic terms to trapezoidal fuzzy values. For example, Highly Important (H) is mapped to trapezoidal fuzzy number (3,5,5,7).

   ii. Compute subjective fuzzy weights of different attributes of good g from the buyer's perspective by combining extent analysis method [5] with fuzzy AHP. Let $\widetilde{\mathrm{FPM}}$ (Fuzzy Pairwise Matrix) represents the fuzzy reciprocal n x n matrix representing all pairwise comparisons $\tilde{a}_{ij}$ for all $i, j \in \{1, 2, \ldots, n\}$ as illustrated in Eq (4) below.

$$\widetilde{\mathrm{FPM}} = \begin{bmatrix} (1,1,1,1) & \tilde{a}_{12} & \cdots & \tilde{a}_{1n} \\ \tilde{a}_{21} & (1,1,1,1) & \cdots & \tilde{a}_{2n} \\ \vdots & \vdots & \ddots & \vdots \\ \tilde{a}_{n1} & \tilde{a}_{n2} & \cdots & (1,1,1,1) \end{bmatrix} \quad (4)$$

Where $\tilde{a}_{ij} = \tilde{a}_{ji}^{-1}$ and all $\tilde{a}_{ij}$ and their inverse $\tilde{a}_{ji}^{-1}$ are trapezoidal fuzzy numbers. The subjective weight computation of attribute $a_i$ denoted as $\widetilde{sw}_{a_i}$ is given in Eq. (5).

$$\widetilde{sw}_{a_i} = \sum_{j=1}^{n} \tilde{a}_{ij} \times \left[\sum_{i=1}^{n} \sum_{j=1}^{n} \tilde{a}_{ij}\right]^{-1} \quad (5)$$

Further, compute $\widetilde{sw}_{a_i}$ for i = 1,2 ..., n, i.e. for all attributes of a good represented by $\widetilde{\mathrm{SW}}$ is shown in Eq. (6).

$$\widetilde{\mathrm{SW}} = \begin{bmatrix} \widetilde{sw}_{a_1} \\ \vdots \\ \widetilde{sw}_{a_n} \end{bmatrix} \quad (6)$$

   iii. Compute the empirical weight component $\widetilde{ew}_{a_i}$, i.e. the average of fuzzy weight of each attribute, for i = 1, 2,..., n, in a maximum of k number of previous transactions by the same buyer for the same good represented by $\widetilde{\mathrm{EW}}$ below.

$$\widetilde{\mathrm{EW}} = \begin{bmatrix} \widetilde{ew}_{a_1} \\ \vdots \\ \widetilde{ew}_{a_n} \end{bmatrix} \quad (7)$$

   iv. Obtain the overall fuzzy attribute weight $\widetilde{w}_{a_i}$ of a good by using Eq. (8) given below.

$$\widetilde{w}_{a_i} = \delta * \widetilde{ew}_{a_i} + (1 - \delta) * \widetilde{sw}_{a_i} \quad (8)$$

Similarly, compute $\widetilde{w}_{a_i}$ for i = 1,2,..,n, represented by $\widetilde{\mathrm{W}}$ as shown in Eq. (9).

$$\widetilde{\mathrm{W}} = \begin{bmatrix} \widetilde{w}_{a_1} \\ \vdots \\ \widetilde{w}_{a_n} \end{bmatrix} \quad (9)$$

In Eq. (8), the value of δ is zero in the case of a buyer purchasing a good for the first time. With each subsequent purchase of the same good by a buyer, the value of δ increases by a small fraction. This ensures that initially when a buyer has no experience of buying a good, the overall weight of a goods' attributes depends only on subjective weight component of each attribute of the good i.e., $\widetilde{sw}_{a_i}$. As buyer gains experience by buying a good repeatedly, the importance of its empirical weight component i.e. $\widetilde{ew}_{a_i}$ increases and the importance of subjective weight component i.e. $\widetilde{sw}_{a_i}$ decreases proportionately. This means that after participating in sufficiently large number of transactions, say k =100 transactions for an δ increment rate of 0.01, by the same buyer for a particular good, it is not necessary for a buyer to incur the overhead of computing the subjective weights of the goods' attributes and instead utilise the previous transactions weight information.

   v. Solicit the buyers' assessment of each seller's offer for the good in linguistic terms like Poor (P), Average (A), High(H), Very High (VH) or Excellent (EX) based on trapezoidal fuzzy scale of Fig. 2.

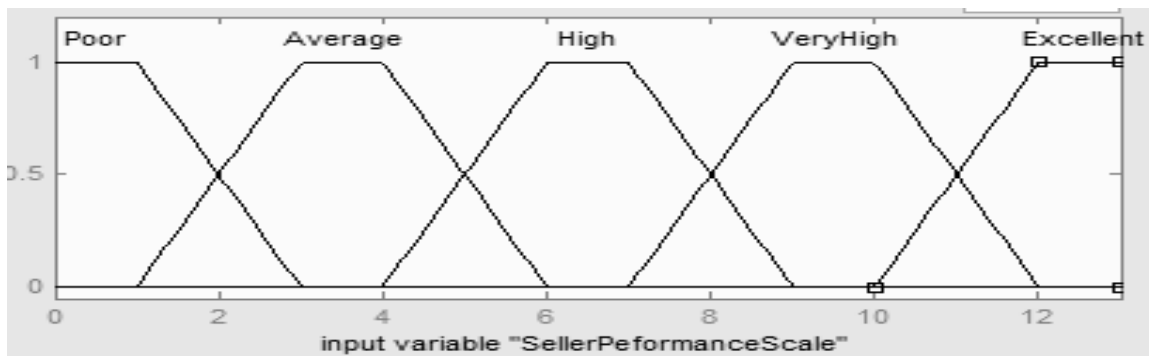

Fig. 2 Fuzzy Scale for Linguistic Performance of Sellers

vi. Using fuzzy scale of Fig. 2, map these linguistic terms into fuzzy performance ratings of good g's offers by different sellers. Let $pr_{ij}$ represents fuzzy performance ratings of seller i for attribute j. Fuzzy performance of each seller i, for i = 1,2..,m and for each attribute j, for j = 1,2,..,n is represented by fuzzy attribute performance matrix $\widetilde{PR}$ in Eq. (10).

$$\widetilde{PR} = \begin{bmatrix} \widetilde{pr}_{11} & \widetilde{pr}_{12} & \cdots & \widetilde{pr}_{1n} \\ \widetilde{pr}_{21} & \widetilde{pr}_{22} & \cdots & \widetilde{pr}_{2n} \\ \vdots & \vdots & \ddots & \vdots \\ \widetilde{pr}_{m1} & \widetilde{pr}_{m2} & \cdots & \widetilde{pr}_{mn} \end{bmatrix} \quad (10)$$

As per Fig. 3, if seller 1's goods' performance for attribute 2 is "VH" then its fuzzy performance rating as per Eq. (10) is $\widetilde{pr}_{12} = (7, 9, 10, 12)$.

vii. Compute the fuzzy value of the seller i's good $\widetilde{fvs}_i$ as: $\widetilde{fvs}_i = \sum_{j=1}^{n} \widetilde{pr}_{ij} \widetilde{w}_{a_j}$ . The fuzzy value matrix of each seller i's good, for i = 1, 2, ..., m represented by $\widetilde{FVS}$ is shown in Eq. (11) below.

$$\widetilde{FVS} = \begin{bmatrix} \widetilde{fvs}_1 \\ \widetilde{fvs}_2 \\ \vdots \\ \widetilde{fvs}_m \end{bmatrix} = \begin{bmatrix} \widetilde{pr}_{11} & \widetilde{pr}_{12} & \cdots & \widetilde{pr}_{1n} \\ \widetilde{pr}_{21} & \widetilde{pr}_{22} & \cdots & \widetilde{pr}_{2n} \\ \vdots & \vdots & \ddots & \vdots \\ \widetilde{pr}_{m1} & \widetilde{pr}_{m2} & \cdots & \widetilde{pr}_{mn} \end{bmatrix} \times \begin{bmatrix} \widetilde{w}_{a_1} \\ \widetilde{w}_{a_2} \\ \vdots \\ \widetilde{w}_{a_n} \end{bmatrix} \quad (11)$$

viii. Perform defuzzification on the fuzzy matrix $\widetilde{FVS}$ to obtain crisp value matrix CVS using Centre of Area approach (COA). CVS contains the crisp expected value i.e. $f^b(g, s)$ of good g's offer from each seller.

ix. Select the seller $s$ with the highest crisp expected value i.e. max. $f^b(g, s)$ of the good $g$ for placing purchase order for the good $g$.

Phase II:

3. Once the buyer receives a good after purchase, it computes the actual value of that good i.e. $v^b(g, s)$, reflecting whether the received good is satisfactory or not as per the buyers' assessment of the actual good by again using step 2 of Phase I.

4. After computing the actual value of a good, buyer updates the individual reputation of seller by first computing the difference between the actual value and the expected value of the good as given in Eq. (12) below.

$$\Delta = v^b(g, s) - f^b(g, s) \quad (12)$$

5. If $\Delta > 0$, then using reinforcement learning technique, buyer b updates reputation of the seller $s$ at time t+1 i.e. $r_{t+1}^b(s)$ with a value greater than its current value as shown in Eq. (13) below.

$$r_{t+1}^b(s) = or_t^b(s) + \mu\,(1 - or_t^b(s)) \quad (13)$$

Where $\mu$ represents effective reputation value increase factor as shown in Eq. (14).

$$\mu = \frac{\eta}{1+\beta} \quad (14)$$

$$\text{and,} \quad \eta = 1 - e^{-\lambda x} \quad (15)$$

Eq. (15) is used to map the value of a transaction $x$ in the range from 0 to 1 which in case of a single good being purchased is equal to the price $p$ of the good $g$. Also $\lambda$ is a constant in the range 0 to 1, and e is a constant with a value of 1.01. The function to compute $\eta$ in Eq. (15) ensures that the value of $\mu$ in Eq. (14) and hence the reputation $r_{t+1}^b(s)$ increases monotonically with the value of transaction. In Eq. (14), $\beta$ is a constant with initial value 0 and its value increases by a small factor, say 0.001, with each successive transaction between the same buyer seller pair. This ensures that with increase in mutual experience of a buyer-seller pair, reputation value i.e. $r_{t+1}^b(s)$ increases at a relatively smaller rate for the same value transaction according to the convention that reputation earned from different buyers is more important than the reputation earned from large number of repeated transactions with the same buyer as shown in Table 1 below.

Table 1: Monotonic Increase of Reputation with value of transaction but discounted with increase in number of transactions between the same buyer-seller pair (For previous reputation i.e. $or_t^b(s) = 0.37$)

| Value of Transaction ($x$) | For $\beta = 0$ | | | For $\beta = 0.5$ | | |
|---|---|---|---|---|---|---|
| | $\mu$ | Updated Reputation | % increase in Reputation | $\mu$ | Updated Reputation | % increase in Reputation |
| 100 | 0.001 | 0.371 | 0.169 | 0.0007 | 0.37 | 0.113 |
| 500 | 0.005 | 0.373 | 0.845 | 0.003 | 0.372 | 0.563 |
| 2000 | 0.02 | 0.3824 | 3.355 | 0.013 | 0.378 | 2.237 |
| 5000 | 0.049 | 0.401 | 8.264 | 0.032 | 0.39 | 5.509 |
| 10000 | 0.095 | 0.4297 | 16.127 | 0.063 | 0.4098 | 10.751 |
| 20000 | 0.18 | 0.4837 | 30.726 | 0.12 | 0.446 | 20.484 |

It can also be observed from Eq. (13) that individual reputation at time t+1 is based on overall reputation at time t to impress upon the fact that in the next transaction, the overall reputation of a seller computed by a buyer at the end of previous transaction becomes the individual experience of that buyer agent. On comparing the relative increase in percentage of reputation in case of a buyer-seller pair having no previous transaction represented by $\beta$ = 0 and after gaining experience of 500 transactions represented by $\beta$ = 0.5, it is found that relative increase in reputation is less in case of $\beta$ = 0.5 as compared to the situation where $\beta$ = 0 as illustrated in Fig. 3.

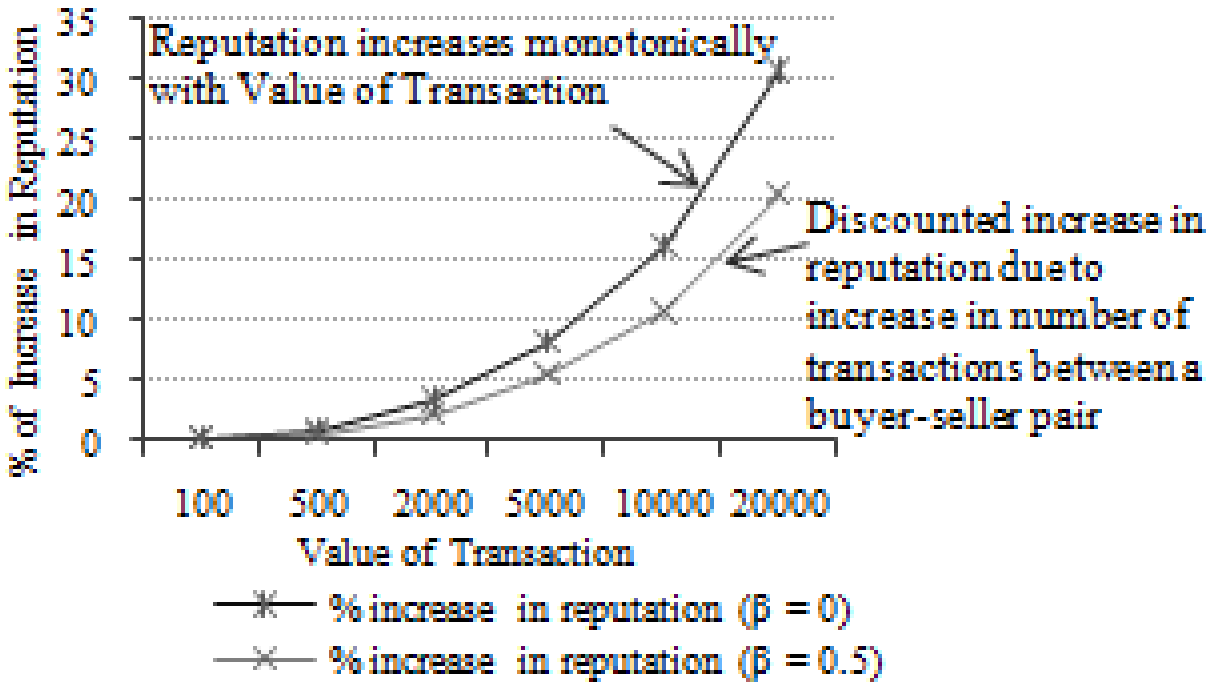

Fig. 3 Monotonic increase of reputation with value of transaction but this increase is discounted/reduced with increase in number of transactions between the same buyer-seller pair to minimise the effect of collusion between a particular buyer and seller

6. If Δ < 0, which represents the fact that the purchased good g has not been satisfactory as per buyer *b*'s assessment, then using reinforcement learning, buyer *b* updates the reputation of the seller *s* at time t+1 i.e. $r_{t+1}^b(s)$ by a value less than its current value as described by Eq. (16).

$$r_{t+1}^b(s) = or_t^b(s) - \xi(1 - or_t^b(s)) \qquad (16)$$

Where ξ represents effective reputation value decrease factor due to unsatisfactory or dishonest behaviour of a seller agent and is illustrated in Eq. (17) below.

$$\xi = \gamma \frac{\eta}{1+\beta} \qquad (17)$$

Where γ is the Penalty Factor and value of γ is kept greater than 1 to ensure that reputation decreases at a faster pace as compared to the rate of its increase. This property is based on the convention that reputation is difficult to build but easy to tear down. The underlying purpose is to discourage dishonest behavior of seller agents in e-market by slapping a higher penalty on fraudulent sellers. Like $\mu$, $\xi$ is also dependent on the value of a transaction and the number of past transactions between a particular buyer-seller pair. Hence there is steep reputation drop for a large value transaction as compared to a small value transaction as described in Table 2.

Table 2: Monotonic Decrease of Reputation with value of a transaction but discounted with increase in number of transactions between the same buyer-seller (Previous reputation, $or_t^b(s)$ = 0.37)

| Value of Transaction ($x$) | For $\beta = 0, \gamma = 2$ | | | For $\beta = 0.5, \gamma = 2$ | | |
|---|---|---|---|---|---|---|
| | $\mu$ | Updated Reputation | % increase in Reputation | $\mu$ | Updated Reputation | % increase in Reputation |
| 100 | 0.002 | 0.3687 | -0.339 | 0.0013 | 0.3692 | -0.226 |
| 500 | 0.01 | 0.3637 | -1.69 | 0.006 | 0.3658 | -1.127 |
| 2000 | 0.039 | 0.3452 | -6.71 | 0.026 | 0.3534 | -4.473 |
| 5000 | 0.097 | 0.3088 | -16.53 | 0.064 | 0.3292 | -11.02 |
| 10000 | 0.189 | 0.2507 | -32.25 | 0.126 | 0.2904 | -21.5 |
| 20000 | 0.361 | 0.1426 | -61.45 | 0.24 | 0.2184 | -40.97 |

The use of the penalty factor γ = 2 applied during reputation computation ensures that the reputation drops at twice the rate as compared to the corresponding rate of its increase for the same value transaction. Comparison of relative increase and decrease in reputation corresponding to the changes in the value of transaction and number of transactions between a particular buyer-seller pair is shown in Fig. 4 below.

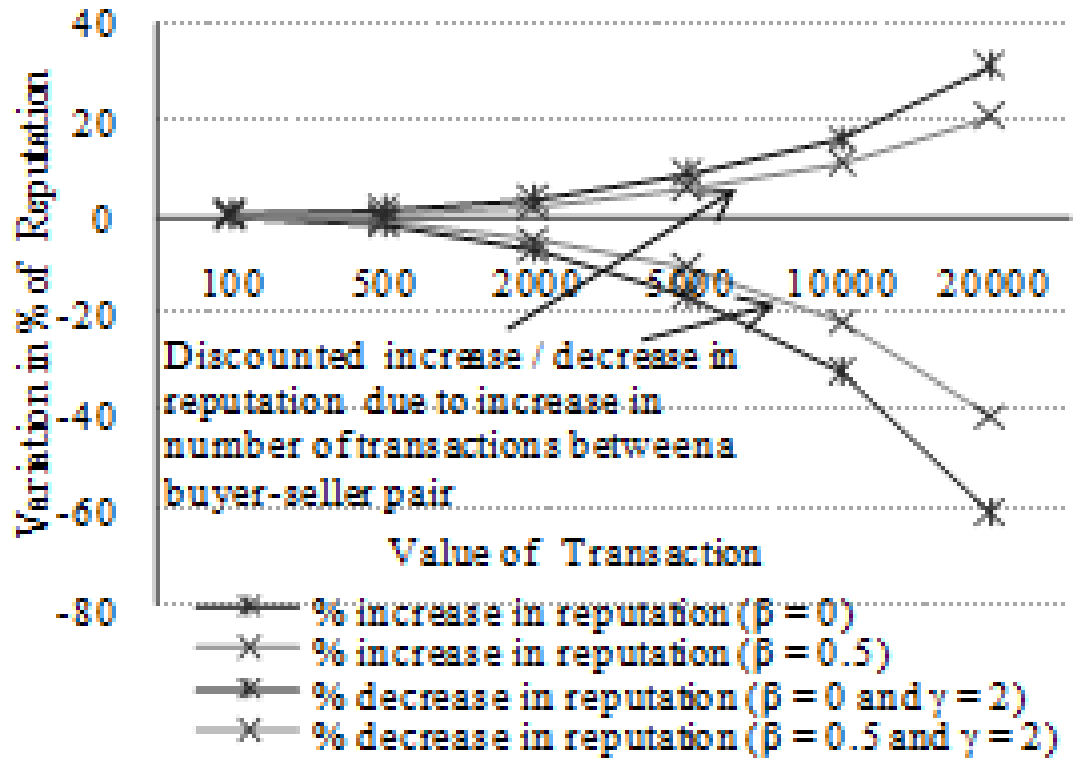

Fig. 4 Reputation drops faster than its increase to discourage dishonest sellers

7. After computing the individual reputation of a seller, this model combines it with the shared reputation about the seller *s* from other buyers to compute the overall reputation of the seller agent *s*. The equation to compute overall reputation function $or_{t+1}^b(s)$ is given below in Eq. (18).

$$or_{t+1}^b(s) = \alpha * r_{t+1}^b(s) + (1 - \alpha) * or_{t+1}^{others}(s) \qquad (18)$$

Where $r_{t+1}^b(s)$ is the individual reputation of seller *s* that is computed by the buyer *b* itself and $or_{t+1}^{others}(s)$ is the aggregate of the reputation rating of seller *s* that is received from other buyer agents. Further, $\alpha$ is the experience gain factor and $0 \le \alpha \le 1$. The initial value of $\alpha$ before the first transaction between a buyer-seller pair is 0 and with each successive transaction, it is incremented by a small factor of say 0.01 to ensure that with each successive transaction between a buyer-seller pair, relative weight of Individual Reputation (IR) component i.e. $r_{t+1}^{b}(s)$ increases and that of Shared Reputation (SR) component i.e. $or_{t+1}^{others}(s)$ decreases. This implements the dynamic property that with increasing mutual transactional experience, a particular buyer-seller pair would depend more on their past mutual experience and less on the opinion from other agents. The actual rate at which the value of $\alpha$ should increase depends on the good to be purchased and is to be decided by domain experts. After sufficiently large number of transactions, as value of $\alpha$ approaches 1, $or_{t+1}^{b}(s)$ would depend only on $r_{t+1}^{b}(s)$ and the weightage of $or_{t+1}^{others}(s)$ would effectively become zero. This ensures that initially when a buyer agent has no experience of a seller, its dependence is greater on the opinion from other buyers although it means incurring some communication overhead. Once a buyer gains sufficient experience of past transactions with a particular buyer, it can avoid the overhead of inter-agent communication as the computation of overall reputation depends only on the individual reputation component. Hence, this framework employs judicious use of information sharing and thus reduces its cost with effective inter-agent communication.

If a seller is new to a buyer $b$ i.e. $s \in S_{NewR}^{b}$ then,

$$or_t^{b}(s) = or_t^{others}(s) \qquad (19)$$

And, if a seller is new in the marketplace, i.e. $s \in S_{NewR}^{\forall b \in B}$ then,

$$or_t^{b}(s) = 0 \qquad (20)$$

Phase III:

8. Finally, on the basis of the overall reputation rating of a seller $s$, sets of reputed, non-reputed, dis-reputed and new sellers i.e. $S_R$, $S_{NR}$, $S_{DR}$ and $S_{NewR}$ are updated as:

If $s$ is not a reputed seller, and $or_{t+1}^{b}(s) \geq \Theta^{b}$ , then

$$S_R^b = S_R^b \cup \{s\}. \qquad (21)$$

If s is a reputed seller, and $or_{t+1}^{b}(s) < \Theta^{b}$ , then

$$S_R^b = S_R^b - \{s\}. \qquad (22)$$

If s is not a dis-reputed seller, and $or_{t+1}^{b}(s) \leq \theta^{b}$, then

$$S_{DR}^b = S_{DR}^b \cup \{s\}. \qquad (23)$$

If $s$ is not non-reputed, and $\theta^{b} < or_{t+1}^{b}(s) < \Theta^{b}$,

$$S_{NR}^b = S_{NR}^b \cup \{s\}. \qquad (24)$$

Finally, if s is a new seller, and, if $or_{t+1}^{b}(s) > \theta^{b}$, then

$$S_{NewR}^b = S_{NewR}^b - \{s\}. \qquad (25)$$

To summarize, the main functions of dynamic reputation framework are illustrated using flowcharts in Fig. 5 and Fig. 6.

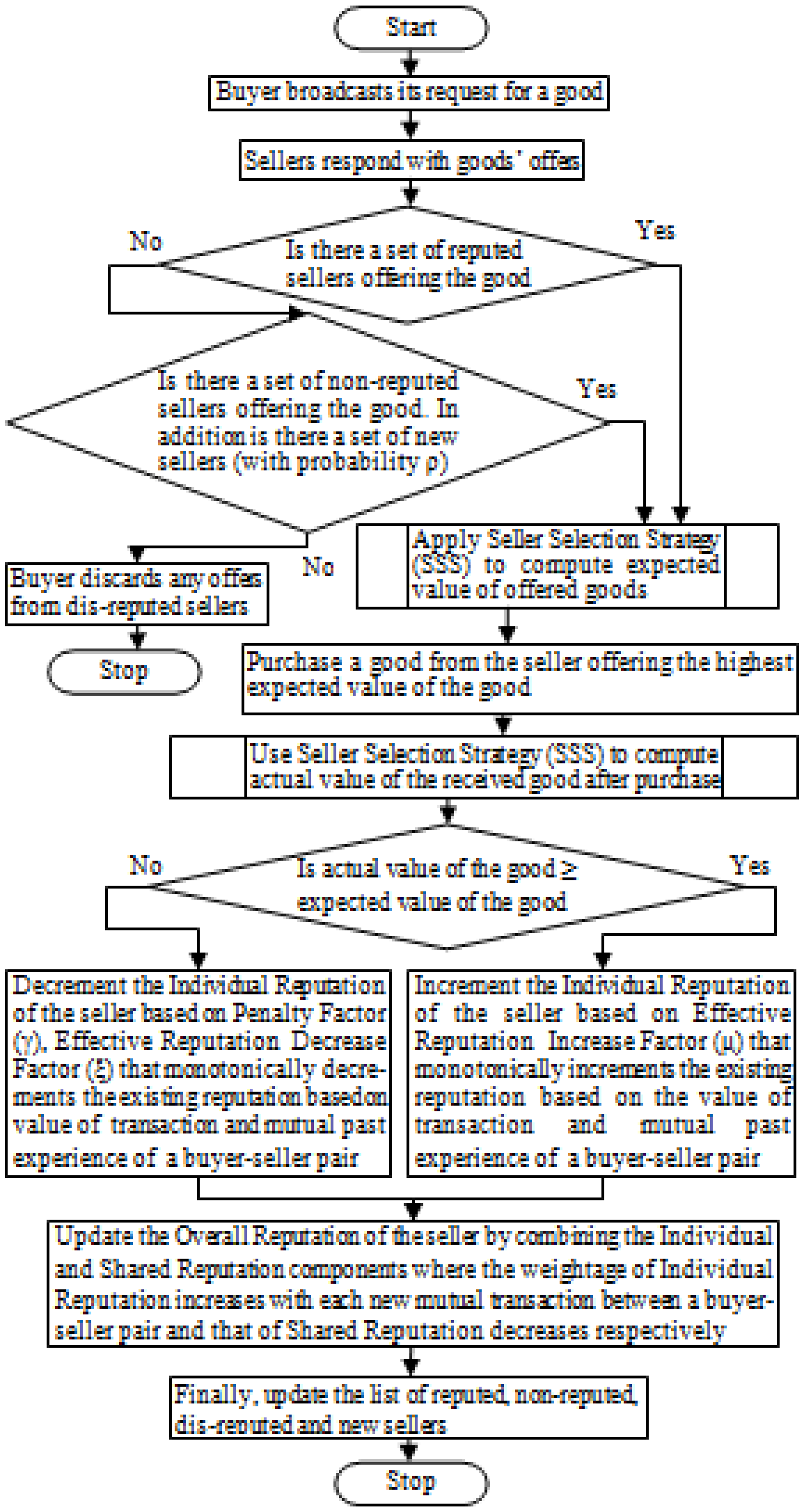

Fig. 5 Dynamic Reputation Framework for Reputation System

The flowchart summarizing the algorithm of seller selection strategy for computing expected/actual value a product is given in Fig. 6 ahead.

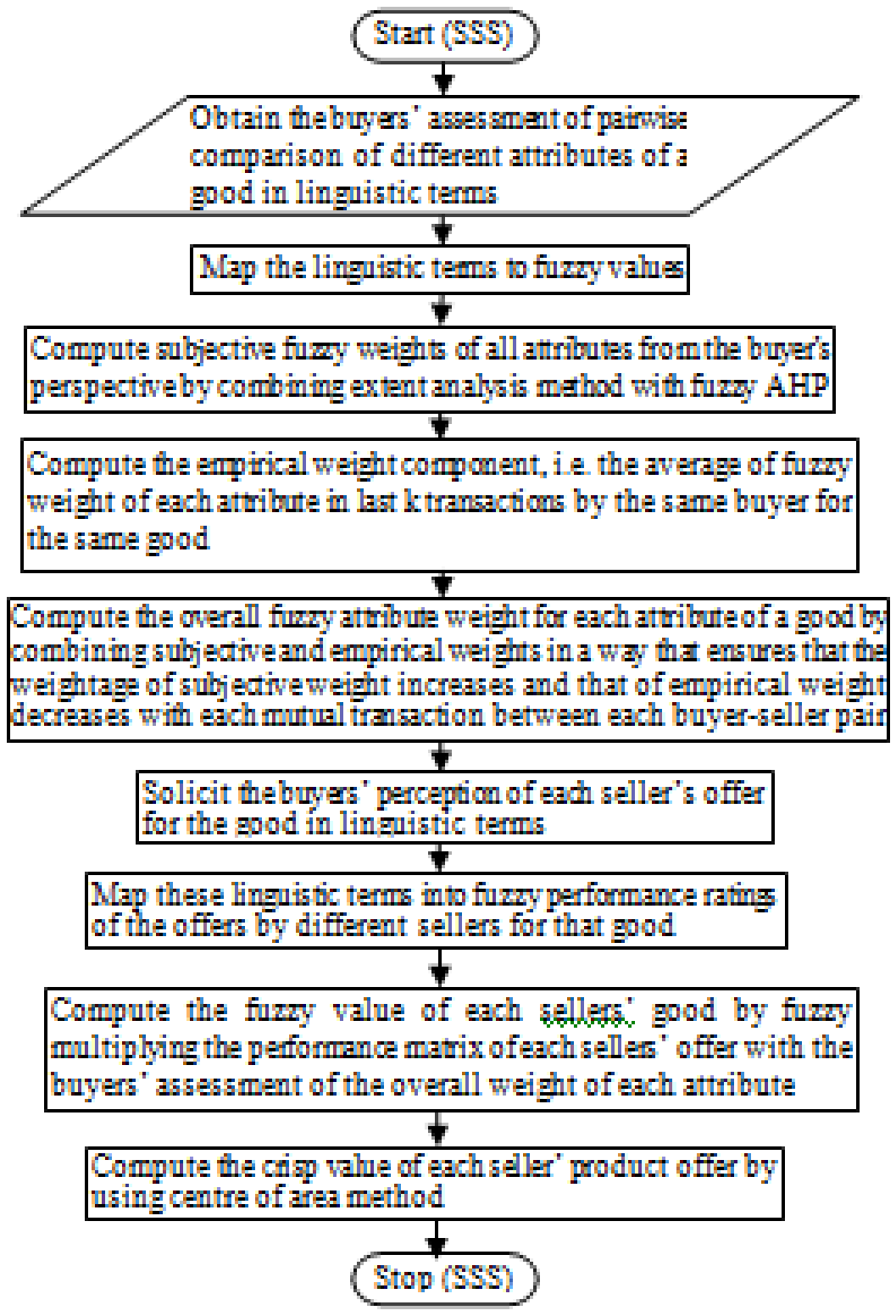

Fig. 6 Seller Selection Strategy (SSS) for computing expected / actual value of a good

## 4. Common Attacks and Proposed Defense

Reputation systems are different from general trust based systems in a way that they include self interested actors or agents who cheat and effectively launch various attacks to defeat these systems. The impact of attacks against reputation systems is much more than the manipulation of reputation values as these result into money fraudulently lost and ruined business reputations. This section discusses different type of attacks classified in literature [2, 3, 18, 22, 24] and presents a comparative performance analysis of the defense capability of the proposed system against these attacks.

In Ballot Stuffing (BS), a group of agents collude to rate a particular agent with abnormally high ratings, whereas in Badmouthing (BM) an agent is rated abnormally low. In this attack, colluding agents participate in events that lead to allocation of reputation or feedback about that agent.

Re-ENtry (REN) is an attack where a low rated agent exits the market and re-enters with a new identity. This attack is facilitated by the availability of cheap pseudonyms in the online environment. The reputation systems with negative feedback are especially vulnerable to REN.

An attack in which two agents mutually rate each other with abnormally high ratings is called RECiprocity (REC) whereas in RETaliation (RET) both the agents rate each other with abnormally low ratings.

Reputation-Lag (RL) takes advantage of the lag i.e. time gap, before cheating results in reduced reputation. During this period, an agent gets unlimited opportunities to cheat before other agents become aware of its loss of reputation due to malicious behaviour.

In Value-IMbalance (VIM) attack, reputation earned or lost during a transaction is not related to value of a transaction. The effect of showing honest behaviour by selling a large number of high quality but low value goods and, dishonest behaviour by selling a small number of low quality but high value goods does not result into any significant loss in reputation score. This helps a malicious seller who behaves honestly for small transactions to gain reputation and then cheats in large transactions.

If a seller agent has no further utility of good reputation, it utilises its entire reputation to cheat and exits from e-market. This attack is called Sudden-Exit (SE).

In Multiple-Identity (MI) or Sybil Attack, a seller is able to open multiple accounts thereby increasing its probability to sell a good. It continues selling the goods honestly through some and dishonestly through others without facing any significant penalty. It exits from the account with a low reputation and opens another account.

Sometimes, a number of attackers employ a combination of strategies to launch a multifaceted and coordinated attack. This is known as Orchestrated (ORC) attack [18]. Attackers change their behaviour overtime and divide themselves into sub-groups where each group plays a different role at different time.

### 4.1 Comparative Performance Analysis

Reputation systems seek to generate an accurate assessment of participants' behaviour in potentially adversarial environments [18]. In uncertain and un-trusted agent based environment of e-market, where the actual buyers and sellers may never meet, absence of such systems may lead to rampant cheating, fraud, mistrust and eventual system failure. Hence, the success of a reputation system is measured by the accuracy of computed reputation that predicts the quality of future interactions in an environment where a participant may try to exploit the system to its own advantage. This section highlights the performance of dynamic reputation framework based on its relative strength as compared to other models from the literature in Table 3.

Table 3: Qualitative analysis of Dynamic Reputation System (DRS) based on known attacks/problems and defense mechanisms

| Type of Attack / Problem | Defense Mechanisms in proposed Dynamic Reputation Framework | Defense Mechanisms in Other Models |
|---|---|---|
| Value Imbalance (VIM) | VIM is resolved as the amount of change in reputation is monotonically related to the value of a transaction. | No model except Truntis deals with this problem. |
| Reputation Lag (RL) | RL is reduced as with increase in the mutual experience of a buyer-seller pair, weightage of shared reputation reduces and finally becomes negligible after large number of transactions. | Models based only on individual reputation [21, 24, 26] do not suffer from RL, other models have no solution. |
| Reciprocity (REC) and Retaliation (RET) | Its effect is minimised as reputation earned by a seller in lieu of repeated transactions with the same buyer is discounted with each successive transaction. Effect of REC/RET is also limited by the value of transaction. | Commercial models like eBay have a strong presence of this attack as 98% of the eBay ratings are positive due to the fear of RET. |
| Re-entry (REN) | REN attack is partially resolved as to re-enter, an agent must lose existing reputation and re-start with minimum reputation. | e-Bay and Truntis deal with this problem with partial success. |
| Sudden Exit (SE) | Probability of SE attack is reduced. As reputation earned is proportional to value of transaction, so to cheat and exit from e- market, an agent has to first earn sufficient reputation by being honest for both large value and large number of transactions. Losing hard earned reputation is not viable unless transaction value is very high. | No feasible solution in any of the proposed model so far. |
| Multiple-Identity (MI) | No inbuilt feasible solution. | No feasible solution provided. |
| Ballot Stuffing (BS)/ Bad-mouthing (BM) | The effect of BS/BM reduces with each successive transaction between a buyer-seller pair as weightage of shared reputation decreases and becomes negligible when an agent gains sufficient experience of other trader agent. | Evidential model, TRAVOS, REGRET and Broker-Assisting TRS try to deal with this attack with varying success. |
| Orchestrated (ORC) | Only partial solution to a subset of attacks is possible as dealing multiple attacks with actors changing roles is very difficult. | No known solution for this type of multifaceted attacks. |

Reputation systems foster good behaviour, punish bad behaviour when it occurs, and reduces the risk of being harmed by others' bad behaviour. Strengths and weaknesses of reputation systems are assessed qualitatively on the basis of their ability to convert the experience of buyer and seller agents into a reputation metric that correctly reflect the behaviour of participants and their capability to withstand different type of attacks launched by dishonest agents. Therefore, a good reputation system must incorporate some inbuilt defense mechanisms to ensure that honest participants are rewarded with economic gains and cheaters are penalised with economic loss. The proposed strategy incorporates inbuilt defense capability in the reputation computation methodology itself by increasing its resilience against various attacks especially Value Imbalance and Ballot Stuffing. It also discourages fraudulent behaviour by slapping a higher penalty on dishonest sellers than the corresponding reward for honest behaviour.

## 5. Case Study

To illustrate the application of proposed reputation framework, a case study was conducted by simulating an electronic marketplace with four users as buyers and six users as sellers, i.e. B = $\{b_i$ where $i = 1...4\}$ and S = $\{s_j$ where $j = 1…6\}$, where B is the set of buyers and S is the set of sellers in the marketplace for good $g$. Some scenarios in the marketplace are shown below.

Scenario 1: A situation was investigated where buyer $b_3$ wanted to buy a good $g$. The sellers $s_1$ to $s_6$ were known to buyer $b_3$, although only three sellers were in its overall reputed list i.e. $S_R^{b3}$ = $\{s_j$ where $j$ = 3,4,5$\}$. Further, $\Theta^{b_3}$=0.45, $\theta^{b_3}$=0.15, e = 1.01, $\alpha$ incremental rate of 0.01 and $\beta$ incremental rate of 0.001 per transaction. Based on buyer $b_3$'s experience, existing overall reputation $or_t^{b_3}(s_j)$ of each seller is depicted in Table 4.

Table 4: Individual reputation ratings of different sellers to buyer $b_3$

| $s_j$ | $s_1$ | $s_2$ | $s_3$ | $s_4$ | $s_5$ | $s_6$ |
|---|---|---|---|---|---|---|
| $or_t^{b3}(s_j)$ | 0.25 | 0.48 | 0.50 | 0.37 | 0.57 | 0.20 |

The buyer $b_3$ specified the pairwise importance of different attributes of good g i.e. of Price (P), Quality (Q), Delivery Period (DP) and Service Offered (SO) in linguistic terms. Their equivalent fuzzy values based on the fuzzy scale of Fig. 2 are shown as Fuzzy Pairwise Matrix ( $\widetilde{FPM}$ ) in (26).

$$\widetilde{FPM} = \begin{array}{c} \\ P \\ Q \\ DP \\ SO \end{array} \begin{array}{c} \begin{array}{cccc} P & Q & DP & SO \end{array} \\ \begin{bmatrix} (1,1,1,1) & (\frac{1}{5},\frac{1}{3},\frac{1}{3},\frac{1}{1}) & (\frac{1}{7},\frac{1}{5},\frac{1}{5},\frac{1}{3}) & (\frac{1}{5},\frac{1}{3},\frac{1}{3},\frac{1}{1}) \\ (1,3,3,5) & (1,1,1,1) & (1,3\ 3,5) & (\frac{1}{5},\frac{1}{3},\frac{1}{3},\frac{1}{1}) \\ (3,5\ 5,7) & (\frac{1}{5},\frac{1}{3},\frac{1}{3},\frac{1}{1}) & (1,1,1,1) & (\frac{1}{5},\frac{1}{3},\frac{1}{3},\frac{1}{1}) \\ (1,3,3,5) & (1,3,3,5) & (1,3,3,5) & (1,1,1,1) \end{bmatrix} \end{array} \quad (26)$$

The average of the weights in the previous transactions were $\widetilde{ew}_P$ = (0.0405,0.115,0.115,0.2435), $\widetilde{ew}_Q$ = (0.11,0.46,0.46,0.87), $\widetilde{ew}_{DP}$ = (0.074,0.196,0.196,0.443)

and, $\widetilde{ew}_{SO}$= (0.0875,0.367,0.367,0.718). Hence, SW , EW, and W of different attributes of the good $g$ as computed in MATLAB with δ = 0.27 are illustrated in Fig. 5 below.

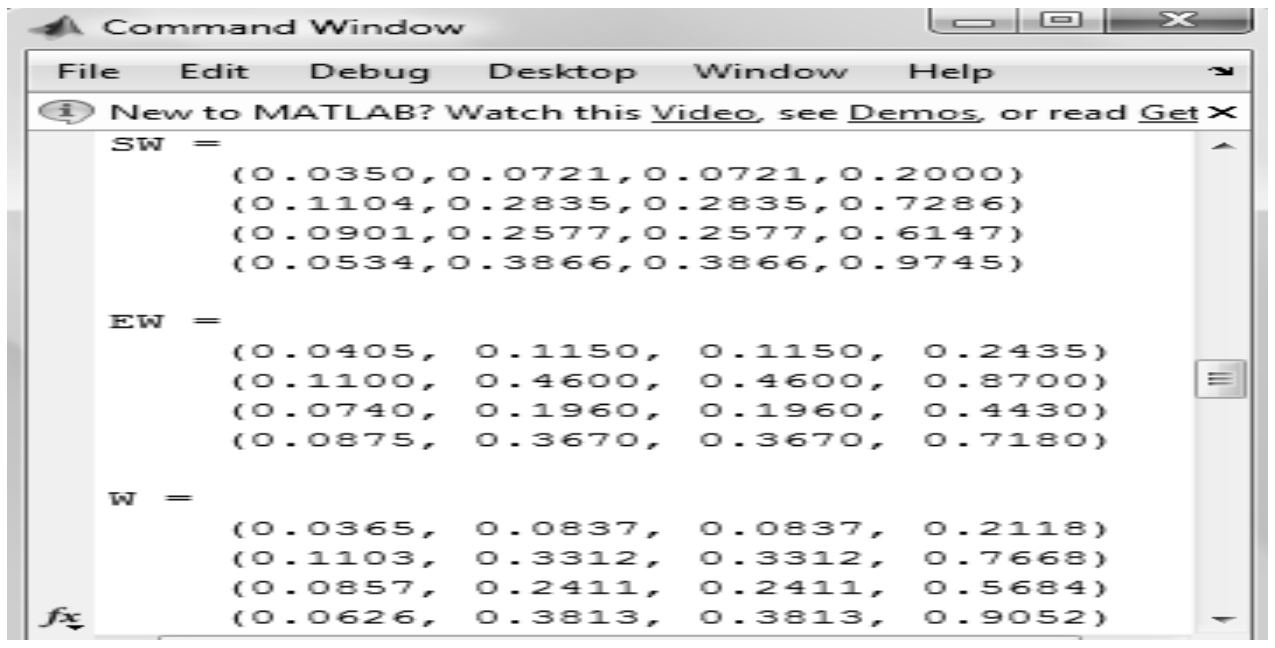

Fig. 5 Overall weight computation of attributes of good $g$ by buyer $b$

Sellers $s_1$, $s_3$, $s_4$, $s_6$, responded to sell good $g$ to buyer $b_1$. Now, buyer $b_1$ computed the expected value of the product being offered by the four sellers as explained below.

After taking buyers' assessment of each seller's product offer for the attributes Price (P), Quality (Q), Delivery Period (DP) and Service Offered (SO) in linguistic terms, the equivalent fuzzy performance matrix $\widetilde{PR}$ representing fuzzy performance of various sellers' offer for good $g$ is shown in (27).

$$\widetilde{PR} = \begin{matrix} & P & Q & DP & SO \\ s_1 & (7,9,10,12) & (7,9,10,12) & (7,9,10,12) & (1,3,4,6) \\ s_3 & (4,6,7,9) & (4,6,7,9) & (7,9,10,12) & (4,6,7,9) \\ s_4 & (4,6,7,9) & (10,12,13,13) & (4,6,7,9) & (4,6,7,9) \\ s_6 & (10,12,13,13) & (7,9,10,12) & (4,6,7,9) & (1,3,4,6) \end{matrix} \quad (27)$$

Using (11), $\widetilde{FVS}$ was computed as, $\widetilde{FVS} = \widetilde{PR} \times \widetilde{W}$ and after defuzzifying the resultant crisp expected value (CVS) representing the expected value of good $g$ for each seller $s_i$ for $i = 1, 3, 4, 6$ , as computed using MATLAB is illustrated in Fig. 6.

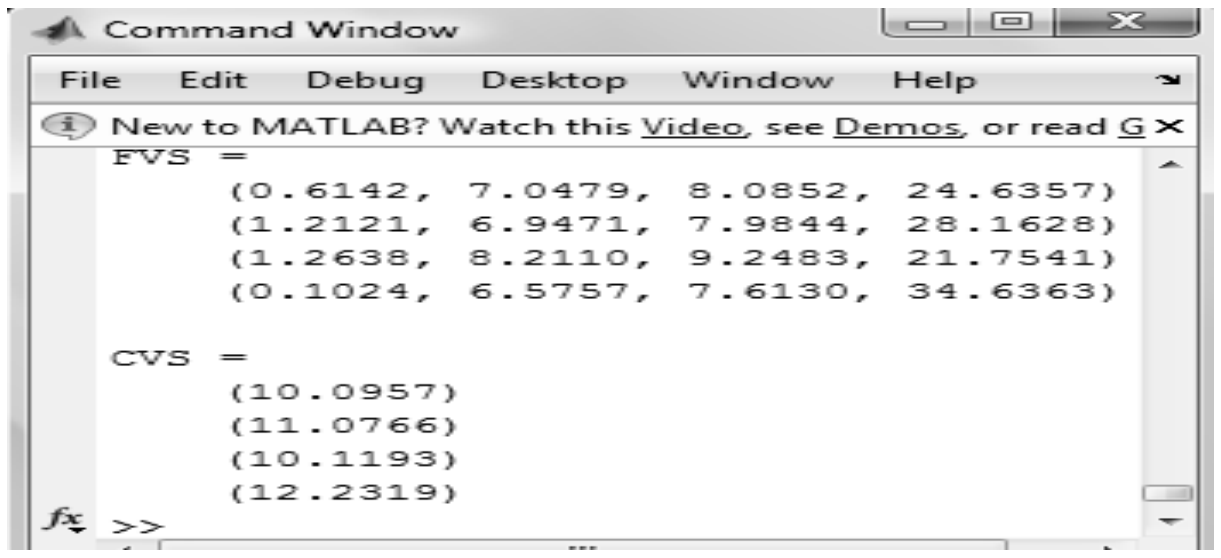

Fig. 6 Fuzzy (FVS) and Crisp (CVS) values of Sellers' offers

Based on Fig. 6, seller $s_6$ with the highest expected value of the good $g$ as 12.2319 is selected by buyer $b_3$ for purchase. Also, as buyer $b_3$ had 79 previous transactions with the seller $s_5$, therefore $\alpha$ = 0.79 and $\beta$ = 0.079. The price of good $g$ by seller $s_5$ was 1500, so $x$ = 1500. After purchasing, and receiving the good $g$, buyer $b_3$ computed the actual value of the good $g$ by again using step 2, Phase I of Section 3 as $v^{b_3}(g, s_5)$ = 13.346.

$$\text{Using (12),} \quad \Delta = 13.146\text{-}12.2319 = 0.9141 > 0. \quad (28)$$

As Δ > 0, buyer $b_3$ incremented the individual reputation of seller $s_5$ as shown below.

$$\eta = 1 - e^{-\lambda x} = 1 - (1.01)^{-0.001*1500} = 0.014815 \quad (29)$$

$$\text{and } \mu = \frac{\eta}{1+\beta} = \frac{0.014815}{1+.079} = 0.01373 \quad (30)$$

$$\text{Using (13), } r_{t+1}^{b_3}(s_5) = 0.57 + 0.01373*(1-0.57) = 0.576. \quad (31)$$

The aggregated shared reputation value for seller $s_5$ was 0.56. Therefore, $b_3$ computed overall rating of seller $s_5$ by combining $r_{t+1}^{b_3}(s_5)$ with $or_{t+1}^{others}(s_5)$ using Eq. (18) as:

$$or_{t+1}^{b_3}(s_5) = 0.79*0.576+(1-0.79)*0.56 = 0.572. \quad (32)$$

Scenario 2: Another situation was investigated where buyer $b_2$ wanted to buy good $g$. Sellers $s_1$ to $s_4$ and $s_6$ were known to buyer $b_2$, whereas sellers $s_3$ and $s_6$ were in its overall reputed list, i.e. $S_R^{b2}$ = {$s_j$ where $j$ = 3,6}. Moreover, $\Theta^{b_2}$=0.5, $\theta^{b_2}$=0.25, $\gamma$ = 3, e = 1.01, $\alpha$ incremental rate of 0.01 per transaction and $\beta$ incremental rate of 0.001 per transaction. After previous transaction of buyer $b_2$, overall reputation ratings $or_t^{b2}(s_j)$ for each seller are given in Table 5.

Table 5: Reputation ratings of different sellers in buyer $b_2$'s memory

| $s_j$ | $s_1$ | $s_2$ | $s_3$ | $s_4$ | $s_6$ |
|---|---|---|---|---|---|
| $or_t^{b_2}(s_i)$ | 0.312 | 0.43 | 0.51 | 0.39 | 0.53 |

Using step 2, Phase I of section 3, the expected value of the good $g$ equivalent to 11.65 was computed to be the maximum for seller $s_3$ so the buyer $b_2$ chose seller $s_3$ to buy good $g$. Also, buyer $b_2$ had 45 previous transactions with the seller $s_3$, therefore $\alpha$ = 0.45 and $\beta$ = 0.045. As seller $s_3$ offered the good $g$ at a price of 6750, so $x$ = 6750. After purchasing, by again using step 2 of Phase I, buyer $b_2$ computed the actual value of good $g$, i.e. $v^{b_3}(g, s_3)$ as 10.87.

$$\text{Using (12),} \quad \Delta = 10.87 - 11.65 = -0.78 < 0. \quad (33)$$

As Δ < 0, buyer $b_2$ decremented the individual reputation of seller $s_5$ as shown below.

$$\eta = 1 - e^{-\lambda x} = 1 - (1.01)^{-0.001*6750} = 0.064959 \quad (34)$$

$$\xi = \gamma \frac{\eta}{1+\beta} = 3 * \frac{0.064959}{1+.045} = 0.18649 \quad (35)$$

$$\text{Using (16), } r_{t+1}^{b_2}(s_3) = 0.51 - 0.18649(1-0.51)) = 0.4186. \quad (36)$$

Further, the aggregated shared reputation value for seller $s_3$ was 0.54. Therefore, $b_2$ finally computed overall rating of seller $s_3$ by combining its individual rating of $s_3$ with $or_{t+1}^{others}(s_3)$ using Eq. (18) as shown below in Eq. (37).

$$or_{t+1}^{b_2}(s_3) = 0.45 * 0.41862 + (1 - 0.45) * 0.54 = 0.4854 \quad (37)$$

Scenario 3: In another case involving Ballot Stuffing attack, buyer $b_4$ needed a good $g$. The sellers $s_1$ to $s_6$ were known to buyer $b_3$ where $s_1$, $s_2$ and $s_4$ are in its reputed list. Further, $\Theta^{b_4}$=0.4, $\theta^{b_4}$=0.18, e = 1.01, α incremental rate of 0.01 and $\beta$ incremental rate of 0.001 per transaction. A number of successive transactions between the buyer $b_4$ and seller $s_2$ were observed where Ballot Stuffing attack was launched on buyer $b_4$ after 20, 50, 75, 95 and 100 transactions between buyer $b_4$ and seller $s_2$. It was seen that the increase in reputation due to BS reduced with the increase in number of transactions as shown in Table 6.

Table 6: Effect of BS reduces with increase in number of transactions between buyer $b_4$ and seller $s_2$

| $or_t^{b_4}(s_2)$ | Number of Transactions | Value of Transaction | $r_{t+1}^{b_4}(s_2)$ | $or_{t+1}^{others}(s_2)$ | $or_{t+1}^{b_4}(s_2)$ | Effect of BS in % Change of Reputation |
|---|---|---|---|---|---|---|
| 0.47 | 20 | 12000 | 0.528 | 0.94 | 0.858 | 62.29 |
| 0.44 | 50 | 1500 | 0.448 | 0.93 | 0.689 | 53.83 |
| 0.48 | 75 | 5300 | 0.505 | 0.95 | 0.616 | 22.01 |
| 0.51 | 95 | 3000 | 0.523 | 0.94 | 0.565 | 3.98 |
| 0.46 | 100 | 2700 | 0.473 | 0.95 | 0.473 | 0 |

It was also observed that the effect of Badmouthing would also be reduced due to reduced weightage of shared reputation with the increase in transactional experience of a buyer-seller pair.

## 6. Conclusions

This paper proposed a framework for a dynamic reputation system that is sensitive to the changing parameters of e-market environment like experience of agents and the value of a transaction in e-market environment. The proposed system has improved inbuilt defense mechanisms against various attacks especially against Ballot Stuffing and Value Imbalance. In this framework, increase in transactional experience leads to increased weightage of individual reputation and honesty in a large transaction leads to a greater increase in reputation as compared to a small transaction. Further, non-satisfactory or fraudulent sellers are penalized with relatively large drop of reputation resulting into reduced future business opportunities. The proposed framework makes judicious use of information sharing by adapting to the changing e-market environment.

**Vibha Gaur:** She is PhD from the department of Computer Science, Delhi University. She is working as Reader in Delhi University and has a teaching experience of about 12 years. She has authored more than 18 papers in various international conferences and journals. Her current research interests include artificial intelligence, information systems and software engineering.

**Neeraj Kumar Sharma:** PhD student at Delhi University. He is also working as Assistant Professor in Delhi University and has a teaching experience of about 8 years. He has published two booklets pertaining to MCA syllabus of IGNOU in the subjects Artificial Intelligence and Algorithms. He has presented two papers in international conferences by ACEEE (ACS 2010) and Springer (ACC 2011). He has also published a paper in International Journal on Recent Trends in Engineering & Technology.